\documentclass[a4paper,10pt,twoside,twocolumn]{article}
\usepackage{macros_ALPHA}
\usepackage[dvips]{epsfig}
\long\def\abstract#1{
        \begin{center}
        \begin{minipage}{\textwidth}
        \normalsize
        \parindent1em\noindent\ignorespaces#1
        \end{minipage}
        \end{center}
        \vskip1.5em}
\def\figurebox#1#2#3{%
        \def\arg{#3}%
        \ifx\arg\empty
        {\hfill\vbox{\hsize#2\hrule\hbox to #2{\vrule\hfill\vbox to
              #1{\hsize#2\
\vfill}\vrule}\hrule}\hfill}%
        \else
        {\hfill\epsfbox{#3}\hfill}%
        \fi}
\begin{document}
%
%%%%%%%%%%%%%%%%%%%%%%%%%%%%%%%%%%%%%%%%%%%%%%%%%%%%%%%%%%%%%%%%%%%%%%%%%%%%%
%%%%%%% Title Page for a Standard Article with Twocolumn Page Layout %%%%%%%%
%%%%%%%%%%%%%%%%%%%%%%%%%%%%%%%%%%%%%%%%%%%%%%%%%%%%%%%%%%%%%%%%%%%%%%%%%%%%%
%
%%%%%%% for workaround to allow for footnote in the title 
%
\renewcommand{\thefootnote}{\fnsymbol{footnote}}
%
%%%%%%% title and article header including preprint number and logo
%
\title{
{\vspace{-2.0cm} \normalsize \hfill
\parbox{21mm}{DESY 00-138}
}\\[20mm]
\textbf{
Applications of non-perturbative renormalization}
\footnotemark[1]
}
\author{
Jochen Heitger
\\[0.25cm]
{\it Deutsches Elektronen-Synchrotron DESY, Platanenallee~6, 
D-15738 Zeuthen, Germany}
\\[0.25cm]
\vbox{
\centerline{
\hspace{-0.25cm}
\epsfxsize=2.5 true cm
\epsfbox{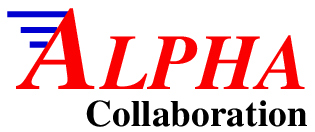}}}
\\[-0.5cm]
}
\date{}
\twocolumn[\maketitle\abstract{{\bf Abstract.}
%
%%%%%%% put in the abstract here
%
A short survey of the renormalization problem in QCD and its
non-perturbative solution by means of numerical simulations on the 
lattice is given.
Most emphasis is on scale dependent renormalizations, which can be 
reliably addressed via a recursive finite-size scaling procedure 
employing a suitable intermediate renormalization scheme. 
To illustrate these concepts we discuss some --- partly recent --- 
computations of phenomenologically relevant quantities: 
the running QCD gauge coupling, renormalization group invariant 
quark masses and the renormalization of the static-light axial
current.
}]
%
%%%%%%% put in title footnote and reset footnote symbols to default
%
\footnotetext[1]{
Invited talk in the session {\it Lattice Field Theory}
at the conference ICHEP '00, July 27 -- August 2, 2000, in Osaka,
Japan.}
\renewcommand{\thefootnote}{\arabic{footnote}}
%
%%%%%%%%%%%%%%%%%%%%%%%%%%%%%%%%%%%%%%%%%%%%%%%%%%%%%%%%%%%%%%%%%%%%%%%%%%%%%

\section{Introduction}
Apart from its well established r\^{o}le as a non-perturbative framework 
to calculate relations between Standard Model parameters and experimental 
quantities from first principles \cite{Kenway:ichep00}, Lattice Field Theory
is particularly designed to solve various renormalization problems in QCD
\cite{schlad:rainer,reviews:leshouches}.
Since renormalized perturbation theory as analytical tool is limited to high
energy processes, where the QCD coupling is sufficiently small, but 
inadequate for bound states and momentum transfers of the order of typical 
hadronic scales, $\mu\simeq 1\,\GeV/c$, a genuinely non-perturbative 
solution of the theory is generally required.
This is achieved by numerical Monte Carlo simulations of the Euclidean QCD
path integral on a space-time lattice.
Though renormalization is an ultraviolet phenomenon (relevant scales 
$\mu^{-1}\sim a$) and QCD asymptotically free, tolerable simulation costs 
prevent the lattice spacing $a$ from becoming much smaller than the 
extent of physical observables so that a truncation of the lattice
perturbative series is often not justified.
Therefore, it is far more safe to perform renormalizations 
non-perturbatively.

In addition, Lattice QCD has a large potential to address the computation of
fundamental parameters of the theory, which escape a direct determination by 
experiments.
The most prominent ones among them are the QCD coupling constant
itself and the quark masses, whose running with the energy scale is
desirable to be understood on a quantitative level beyond perturbation 
theory --- the central subject of the next sections.
The knowledge of these quantities (e.g.~at some common reference point, 
$\alphs(\MZ)$ or $\mbar(2\,\GeV)$) might then also provide essential
input to theoretical analyses of observables of phenomenological interest. 
For instance, the mixing ratio $\epsilon'/\epsilon$ in the neutral kaon 
system incorporates the strange quark mass value.

The renormalization properties of many other quantities have been 
investigated with lattice methods, e.g.~(bilinear) quark composite 
operators, $\Delta S=2$ matrix elements and, as presented at this 
conference too, structure functions \cite{Jansen:ichep00}.
Later we will briefly discuss the non-perturbative renormalization of the 
static axial current as a further example.

During the last few years the lattice community has seen much theoretical 
and numerical advances \cite{Kenway:ichep00,proceedings:lat99}.
Here it is worth to mention at least the issue of $\Or(a)$ discretization
effects inherent in the Wilson fermion action.
In case of the quenched approximation to QCD, where all dynamics due to 
virtual quark loops is ignored, they have been systematically eliminated 
through a non-perturbative realization of Symanzik's improvement programme 
\cite{schlad:rainer,reviews:leshouches,tsuk97:rainer}.
Hence, lattice artifacts can be extrapolated away linearly in $a^2$, which 
allows to precisely extract many physical quantities in the continuum 
limit, $a\rightarrow 0$.

\section{Intermediate schemes}
%
%%% Beginn Tabelle %%%
\begin{table*}[htb]
\begin{center}
\begin{tabular}{rcccp{187pt}}
$\lmax=C/\Fpi$: $\Or(\half\mbox{fm})$ 
& hadronic scheme $\hookrightarrow$ SF 
& $\longrightarrow$ & $\alphSF$\,($\mu=1/\lmax$)       \\
& & & $\downarrow$                                     \\
& & & $\alphSF$\,($\mu=2/\lmax$)                       \\
& & & $\downarrow$                                     \\
& & & $\bullet\bullet\bullet$                          \\
& & & $\downarrow$                                     \\
& & & $\alphSF$\,($\mu=2^n/\lmax$)                     \\
& & & $\downarrow$                                     \\
& & & PT                                               \\
& & & $\downarrow$                                     \\
jet physics ($e^+e^-\rightarrow q\,\overline{q}\,g$) 
$\hookleftarrow$ & value for $\lQCD/\Fpi$ 
& $\stackrel{\mbox{PT}}{\longleftarrow}$ & $\lSF\lmax$ \\
\end{tabular}
\end{center}
\end{table*}
%%% Ende Tabelle %%%
%
As a representative example for a non-perturbative renormalization problem
we may consider the calculation of quark masses through the PCAC relation,
\bea
\Fk\mk^2
& = &
(\mUb+\msb)\ketbra{0}{\overline{u}\gfv s}{{\rm K}}
\label{pcac_k}\\
(\overline{u}\gfv s)_{\MS}
& = &
\zp(g_0,a\mu)(\overline{u}\gfv s)_{\lat}\,,
\label{zp_def}
\eea
in which the scale and scheme dependent renormalization constant $\zp$ 
relates the lattice results to the $\MS$ scheme and is computable in 
lattice perturbation theory.
But since this expansion introduces errors which are difficult to control, 
a non-perturbative determination of the renormalization factor is needed.
A non-perturbative renormalization condition between the two schemes can,
however, not be formulated, because $\MS$ is only defined perturbatively.

The idea to overcome this problem is the introduction of an intermediate
renormalization scheme: the lattice observable is first matched at some 
fixed scale $\mu_0$ to the corresponding one in the intermediate scheme,
and afterwards it is evolved from $\mu_0$ up to high energies, where 
perturbation theory (PT) is expected to work well.
Nonetheless, as in a simulation one then has to cover many scales (the box 
size $L$, $\mu\simeq0.2\,\GeV\,-\,10\,\GeV$ and the lattice cutoff $a^{-1}$) 
simultaneously, the task to reliably match the low energy regime with the 
high energy one, i.e.~the applicability domain of perturbation theory, gets 
quite complicated.
In the present context two implementations of such schemes are available, 
the regularization independent approach \cite{RI:Rome} and the QCD 
Schr\"odinger functional (SF) \cite{SF:LNWW,SF:stefan1}.
Whereas the former may suffer from the scale hierarchy problem in practice, 
the basic strategy of the SF approach is to recourse to an intermediate 
finite-volume renormalization scheme, where one identifies two of the 
before-mentioned scales, $\mu=1/L$, and takes low energy data as input in 
order to use the non-perturbative renormalization group to scale up to high 
energies \cite{schlad:rainer,reviews:leshouches}.

A schematic view of a non-perturbative computation of short distance 
parameters on the lattice along these lines, here in case of the running 
QCD coupling $\alpha(\mu)$, is given in the diagram above; the same can 
also be set up for the running quark masses.
It is important to note that all relations `$\rightarrow$' are accessible in 
the continuum limit and in this sense universal by construction.

\section{$\lQCD$ and $M_{\rm quark,\,RGI}$ via the SF}
The Schr\"odinger functional is the QCD partition function with certain 
Dirichlet boundary conditions in time imposed on the quark and gluon fields,
for which a renormalized coupling constant can be defined as the response to 
an infinitesimal variation of the boundary conditions 
\cite{SF:LNWW}.
By help of the so-called step scaling function, being a measure for the 
change in the coupling when changing the box size $L$ (and thus having the
meaning of a discrete $\beta$--function), one is now able in the SF scheme 
to make contact with the high-energy regime of perturbative scaling:
\bea
\Lambda
& \equiv &
\lim_{\mu\to\infty}\left\{
\mu(b_0\gbsq(\mu))^{-b_1/2b_0^2}\,\Exp^{\,-1/2b_0\gbsq}\right\} 
\nonumber\\
&        &
b_0=11/(4\pi)^2\,,\,b_1=102/(4\pi)^4\,.
\label{lambda_def}
\eea
Every step during the non-perturbative evolution towards the perturbative
regime has been extrapolated to the continuum limit in the quenched 
approximation \cite{mbar:pap1}, and upon conversion to the $\MS$ scheme 
this results in a value for the $\Lambda$--parameter:
\be
\lMSbarq=238(19)\,\MeV\,.
\label{LambdaQCD_res}
\ee
An extension of this investigation to the situation with two dynamical 
quarks is already in progress by the ALPHA Collaboration.

\begin{figure}[htb]
\begin{center}
\epsfxsize 8.0cm
\figurebox{8.0cm}{}{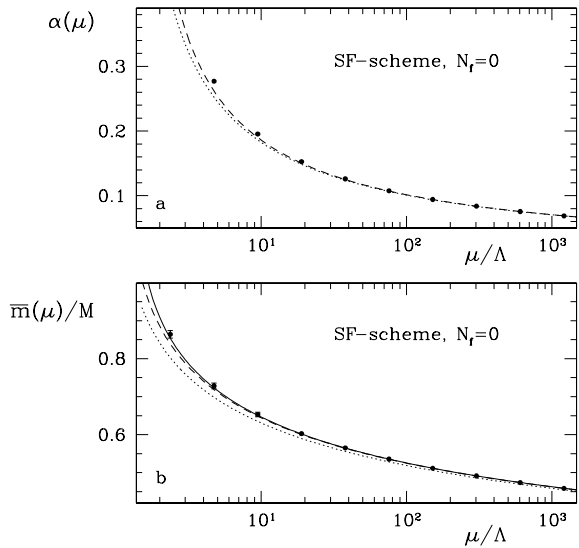}
\vspace{-0.5cm}
\caption{Non-perturbative scale evolution of $\alphSF$ and $\mbSF/M$
computed from simulations of the SF in the quenched approximation.
The lines represent perturbative predictions involving the 2-- and
3--loop $\beta$--function (a) and 1/2--, 2/2-- and 2/3--loop expressions
for the $\tau$-- and $\beta$--functions, respectively (b).}
\label{AlphMbarPlot}
\end{center}
\vspace{-0.5cm}
\end{figure}

In a very similar way, in terms of the current quark mass renormalization 
factor $\zp$ of eq.~(\ref{zp_def}) replacing the SF coupling to build up 
another step scaling function, the scale and scheme independent
renormalization group invariant (RGI) quark masses
\bea
M 
& \equiv &
\lim_{\mu\to\infty}\left\{
(2b_0\gbsq(\mu))^{-d_0/2b_0}\,\mbar(\mu)\right\} \nonumber\\
&        &
b_0=11/(4\pi)^2\,,\,d_0=8/(4\pi)^2
\label{mRGI_def}
\eea
were obtained in the same reference.
Both evolutions are displayed in Fig.~\ref{AlphMbarPlot}, and at the 
scale $\mu_0$ (leftmost point in Fig.~\ref{AlphMbarPlot}b)
the matching between the lattice regularization and $\MS$ via the SF is
completed:
\be
\frac{M}{\mbSF(\mu_0)}=1.157(12)\,,\,
\mu_0\simeq 275\,\MeV\,.
\label{MmbarSF_res}
\ee
%
%%% Beginn Figur %%%
%%% Ende Figur %%%
%

For the $\Or(a)$ improved theory and a massless renormalization scheme as 
utilized here, these results can also be summarized as
\bea
M
& = &
\zM(g_0)\times m(g_0)+\Or(a^2) \nonumber\\
\zM(g_0)
& = &
\frac{M}{\mbar(\mu)}\times\frac{\mbar(\mu)}{m(g_0)}\,,
\label{ZM_def}
\eea
where $m$ is the bare current quark mass, and the flavour independent total 
renormalization factor $\zM$, non-perturbatively known for a range of bare 
couplings $g_0$ in the quenched approximation \cite{mbar:pap1}, is composed
of an universal part, $M/\mbar$, and of $\mbar/m=\za/\zp$ depending on the 
lattice regularization.

%%% Local Variables: 
%%% mode: latex
%%% TeX-master: "sect4"
%%% End: 

\section{The strange quark's mass}
In order to illustrate the non-perturbative quark mass renormalization
just explained in a concrete numerical application, we first sketch our 
strategy for the computation of light quark masses \cite{mbar:pap3}.
Their ratios are known from chiral perturbation theory ($\cpt)$ 
\cite{reviews:quarkmasses} as
\be
\frac{\Mu}{\Md}=0.55(4)\,,\quad 
\frac{\Ms}{\Mhat}=24.4(1.5)  
\label{m_ratios}
\ee
with $\Mhat=\half(\Mu+\Md)$ \cite{leutwyler:1996}.
Nevertheless there are still questions, which might be answered decisively
only using Lattice QCD.
They concern the applicability of $\cpt$ in general, i.e.~in how far the 
lowest orders dominate the full result, and the problem that the parameters
in the chiral Lagrangian (at a given order in the expansion) can not be 
inferred with great precision from experimental data alone.
This statement holds in particular for the overall scale of the quark 
masses, which is only defined once the connection with the fundamental 
theory, QCD, is made. 
Since the parameters in the chiral Lagrangian (the so-called low energy 
constants) are independent of the quark masses, it is important to realize 
that these problems can be dealt with by working with unphysical --- of 
course not too large --- quark masses, where it is essential or at least 
of significant advantage to explore a certain range of quark masses.
While a determination of some low energy constants based on these ideas has
been recently tested in \cite{mbar:pap4}, we focus in the following on the 
computation of the renormalization-group invariant mass of the strange 
quark by combining $\cpt$ with lattice techniques.

To this end, and in the spirit of the considerations before, we define a 
reference quark mass $\Mref$ implicitly through
\bea
\mps^2(\Mref)r_0^2=(\mk r_0)^2=1.5736\,.
\label{mref_def}
\eea
Here $\mps^2(M)$ is the pseudoscalar meson mass as a function
of the quark mass for mass-degenerate quarks, and $r_0=0.5\,\Fm$ and
$\half(\mkp^2+\mkz^2)\big|_{\rm pure\,\,QCD}=(495\,\MeV)^2$ enter the r.h.s. 
of eq.~(\ref{mref_def}).
$\cpt$ in full QCD relates $\Mref$ to the other light quark masses viz. 
\bea
2\Mref \simeq \Ms+\Mhat\,,
\label{mref_ms}
\eea
which has been substantiated also numerically in the case of quenched QCD 
\cite{mbar:pap3}.
The remaining task is now to calculate $\Mref$ from Lattice QCD.

\begin{figure}[htb]
\begin{center}
\epsfxsize 8.0cm
\figurebox{8.0cm}{}{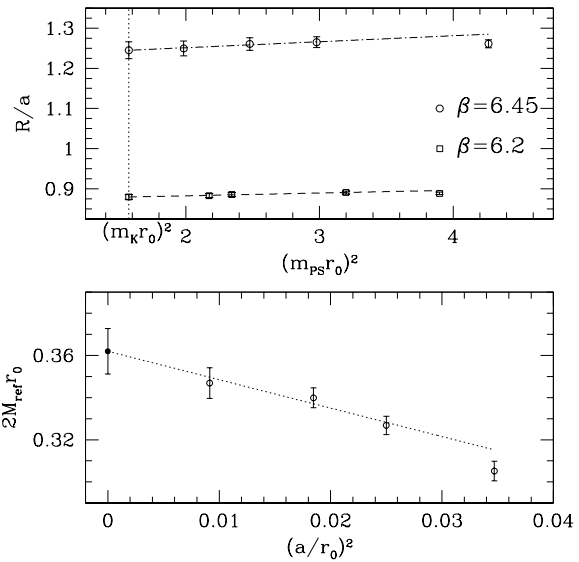}
\caption{Extrapolations of the ratio $R$ to the kaon mass scale and, in
units of $r_0$, to the continuum limit.}
\label{ExtrPlot}
\end{center}
\end{figure}

As the foregoing discussion holds true in mass independent renormalization
schemes too, one arrives by virtue of the PCAC relation applied to the 
vacuum-to-pseudoscalar matrix elements at the central relation
\bea
2r_0\Mref
&    =   & 
\zM\frac{R\,|_{\mps^2r_0^2=1.5736}}{r_0}\,1.5736 \nonumber \\
R           
& \equiv & 
\frac{\Fps}{\Gps}\,,
\label{Mr0_def}
\eea
where $\zM$ is the flavour independent renormalization factor of the 
previous section, which directly leads to the RGI quark masses, being pure 
numbers and not depending on the scheme.
By means of numerical simulations of the SF in large volumes of size
$(1.5\,\Fm)^3\times 3\,\Fm$, the ratio $R/a$ and the meson mass $\mps a$ 
can be computed accurately as a function of the bare quark mass and the 
bare coupling by evaluating suitable correlation functions 
\cite{mbar:pap2,mbar:pap3}.
With the values for the scale $r_0/a$ from \cite{pot:r0_SU3}, a mild 
extrapolation yields $R/a$ at the point $\mps^2r_0^2=1.5736$.
Then the quantity $2r_0\Mref$ is extrapolated to the continuum limit.
Both fits are shown in Fig.~\ref{ExtrPlot}.
In view of the still significant slope in the latter, we discard the point 
furthest away from the continuum in this extrapolation as a safeguard 
against higher order lattice spacing effects.
Moreover, the analysis was repeated for $\Mref$ in units of the kaon decay
constant, which amounts to substitute eq.~(\ref{Mr0_def}) by
\be
\frac{2\Mref}{(\Fk)_{\rm R}}=
\frac{M}{\mbar}\,\frac{1}{\zp\,r_0^2\,\Gps}\,1.5736\,.
\label{MFk_def}
\ee
Here we observe a weaker lattice spacing dependence.
The final results of these analyses
\bea  
2r_0\Mref=0.36(1) \nonumber \\
\stackrel{r_0=0.5\,\Fm}{\longrightarrow} 
& 2\Mref=143(5)\,\MeV & \nonumber \\
\frac{2\Mref}{(\Fk)_{\rm R}}=0.87(3) \nonumber\\
\stackrel{(\Fk)_{\rm R}=160\,\MeV}{\longrightarrow}  
& 2\Mref=140(5)\,\MeV & \nonumber 
\eea
are completely consistent with each other.
But, as also pointed out in that reference, the assignment of physical units 
is intrinsically ambiguous in the quenched approximation. 
Consulting e.g.~the recent results of the CP-PACS Collaboration 
\cite{qspect:CPPACS}, roughly $10\,\%$ larger numbers would be obtained,
if the scale $r_0$ were replaced by one of the masses of the stable light
hadrons.
$\MS$ masses for finite renormalization scales $\mu$ can be obtained through
perturbative conversion factors known up to 4--loop precision.
A typical result is
\be
\msbMS(2\,\GeV)=97(4)\,\MeV\,,
\label{ms_2GeV}
\ee
where the uncertainty in $\lMSbarq$, eq.~(\ref{LambdaQCD_res}), entering 
the relation of the running quark masses in the $\MS$ scheme to the RGI 
masses, eq.~(\ref{mRGI_def}), and the quark mass ratios from full QCD chiral
perturbation theory, eq.~(\ref{m_ratios}), were taken into account
\cite{mbar:pap3}.
%
%%% Beginn Figur %%%
%%% Ende Figur %%%
%

A compilation of lattice results on the strange quark mass in the quenched 
approximation can be found e.g.~in \cite{Wittig:epshep99}.
Most of these differ in the Ward identity used and in whether 
non-perturbative renormalization and a continuum extrapolation has been 
performed or not; also systematic errors often are not estimated 
uniformly either.
Our result (\ref{ms_2GeV}) includes all errors except quenching.
Finally it is interesting to note that, as reported by the 
CP-PACS Collaboration in their comprehensive study about simulations with 
two dynamical flavours \cite{mquark_nf2:CPPACS}, dynamical quark effect 
appear to decrease the estimates for the strange quark mass by $\sim20\,\%$
or less.

%%% Local Variables: 
%%% mode: latex
%%% TeX-master: "sect5"
%%% End: 

\section{The static-light axial current}
Let us turn to another example, where a scale and scheme dependent 
renormalization is encountered, i.e.~the matrix element 
$\ketbra{0}{\Amur}{{\rm B}(p)}=ip_{\mu}\Fb$ describing leptonic B--decays
in the theory with heavy quarks.
It involves the renormalized axial current,
$\Amur=\za\overline{b}\gmu\gfv d$, and the decay constant $\Fb$, which is 
by its own an interesting quantity for a first principles computation
on the lattice. 
Since $\mb\simeq4\,\GeV\gg\lQCD$ implies large discretization errors of
$\Or\left((a\mb)^2\right)$, a direct treatment assuming a relativistic
b quark is difficult on the lattice.
Therefore, in the first place one may restrict to an effective theory, one 
possibility being the static approximation, where the b quark is taken
to be infinitely heavy.

%
%%% Beginn Figur %%%
\begin{figure}[htb]
\begin{center}
\vspace{-0.75cm}
\epsfxsize 8.0cm
\figurebox{8.0cm}{}{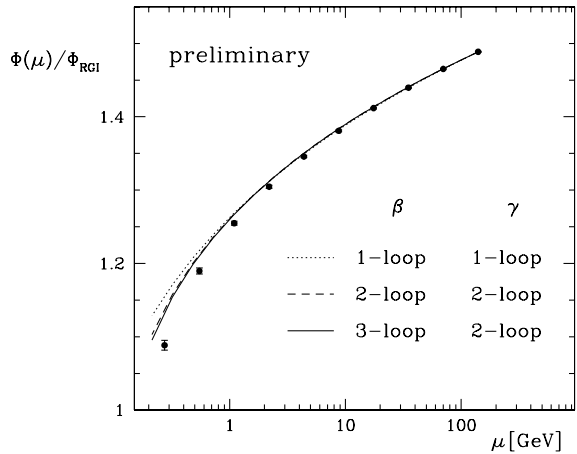}
\vspace{-0.75cm}
\caption{Non-perturbative running of $\Phi/\Phi_{\rm RGI}$ with the
energy scale in the static approximation, computed in the SF scheme, 
compared to perturbation theory based on several combinations of orders,
to which the $\beta$-- and $\gamma$--functions have been evaluated.}
\label{PhiPlot}
\end{center}
\end{figure}
%%% Ende Figur %%%
%
As at the end we want to relate the physical matrix element $\Phi$ at a 
scale $\mu=\mb$,
\be 
\Fb\sqrt{\mB}\equiv\Phi(\mu)+\Or\left(\frac{\lQCD}{\mb}\right)\,,
\label{phi_def}
\ee
to the one determined on the lattice at some matching scale $\mu_0$, a 
crucial ingredient is its (scale and scheme independent) renormalization 
group invariant counterpart
\bea
&   &
\Phi_{\rm RGI}\equiv
\lim_{\mu\to\infty}\left\{
(2b_0\gbsq(\mu))^{-\gamma_0/2b_0}\,\Phi(\mu)\right\} \nonumber\\
&   &
b_0=11/(4\pi)^2\,,\,\gamma_0=-1/4\pi^2
\label{phiRGI_def}
\eea
to be passed into the factorization
\be
\Phi(\mu)=
\frac{\Phi(\mu)}{\Phi_{\rm RGI}}\,
\frac{\Phi_{\rm RGI}}{\Phi_{\rm SF}(\mu_0)}\,
\Phi_{\rm SF}(\mu_0)\,.
\label{phi_match}
\ee
As already anticipated in the notation, the further strategy is basically
analogous to that explained when considering the coupling and the quark 
masses: we again adopt the SF framework and invoke an appropriate step 
scaling function, while everything is meant in the static approximation now.

The definition of the renormalized static axial current and the step 
scaling function, together with the 2--loop anomalous dimension, has 
recently been worked out perturbatively \cite{zastat:lat99pap1}.
The preliminary status of the outcome of the corresponding non-perturbative
investigation by numerical simulations of the SF in the quenched 
approximation is depicted in Fig.~\ref{PhiPlot}.
The leftmost factor in eq.~(\ref{phi_match}) is then supposed to be 
inferable in that region, where perturbation theory is feasible again.

\section{Conclusions}
Numerical simulations on the lattice can be applied to renormalization 
problems in QCD.
In particular, the Schr\"odinger functional scheme offers a clean and 
flexible approach to deal with the accompanying scale differences.
As a consequence of good control over statistical, discretization and 
systematic errors, non-perturbative coupling and quark mass 
renormalization can be performed with confidence, and solid results for
$\lMSbarq$ and $\msbMS$ with high precision of the order of a few \%
were reached in the quenched approximation.
Similar ideas are now carried over to the heavy quark sector of QCD, 
where first steps towards a computation of renormalization group invariant
matrix elements in the static approximation are under way.

The presented concepts will be valuable also for full QCD.
Despite more powerful (super-)computers continuously being developed, a
quantitative understanding of dynamical sea quark effects is a great 
challenge which, albeit in sight, still demands for much effort on the 
theoretical as well as on the technical/implementational side of 
Lattice QCD.

\end{document}